\documentclass[a4paper,10pt]{article}

\usepackage{epsf}
\usepackage{graphicx}    

\newcommand{\be}[1]{\begin{equation}\label{#1}}
\newcommand{\ee}{\end{equation}}
\newcommand{\bea}{\begin{eqnarray}}
\newcommand{\eea}{\end{eqnarray}}
\def\disp{\displaystyle}

\def\gsim{ \lower .75ex \hbox{$\sim$} \llap{\raise .27ex \hbox{$>$}} }
\def\lsim{ \lower .75ex \hbox{$\sim$} \llap{\raise .27ex \hbox{$<$}} }

\renewcommand{\markright}{\markright{\thepage}}

\begin{document}

\begin{titlepage}

\begin{flushright}
arXiv:0704.3330
\end{flushright}

\vspace{5mm}

\begin{center}

{\Large \bf Interacting Energy Components
 and Observational $H(z)$ Data}

\vspace{10mm}

{\large Hao~Wei$^{\,1,}$\footnote{\,email address:
 haowei@mail.tsinghua.edu.cn} and Shuang~Nan~Zhang$^{\,1,2,3}$}

\vspace{5mm}
{\em $^1$Department of Physics and Tsinghua Center for
 Astrophysics,\\ Tsinghua University, Beijing 100084, China\\
 $^2$Key Laboratory of Particle Astrophysics, Institute of High
 Energy Physics,\\ Chinese Academy of Sciences, Beijing 100049,
 China\\
 $^3$Physics Department, University of Alabama in Huntsville,
 Huntsville, AL 35899, USA}

\end{center}

\vspace{5mm}
\begin{abstract}
In this note, we extend our previous work [Phys.\ Lett.\  B
 {\bf 644}, 7 (2007), astro-ph/0609597], and compare
 eleven interacting dark energy models with different couplings
 to the observational $H(z)$ data. However, none of these
 models is better than the simplest $\Lambda$CDM model.
 This implies that either more exotic couplings are needed in the
 cosmological models with interaction between dark energy and dust
 matter, or {\em there is no interaction at all}. We consider that
 this result is disadvantageous to the interacting dark energy
 models studied extensively in the literature.\\

\noindent PACS numbers: 95.36.+x, 98.80.Es, 98.80.-k

\end{abstract}

\end{titlepage}

\newpage

\setcounter{page}{2}

\section{Introduction}\label{sec1}

Nowadays, dark energy study has been one of the most active fields
 in modern cosmology~\cite{r1}, since the discovery of the present
 accelerated expansion of our universe~\cite{r2,r3,r4,r5,r6,r7}.
 In the past years, many cosmological models are proposed to
 interpret this phenomenon. One of the important tasks is to
 confront them with observational data. The most frequent
 method to constrain the model parameters is fitting them to the
 luminosity distance
 \be{eq1}
 d_L(z)=(1+z)\int_0^z \frac{d\tilde{z}}{H(\tilde{z})}\,,
 \ee
 which is an integral of Hubble parameter $H\equiv\dot{a}/a$, where
 $a=(1+z)^{-1}$ is the scale factor ($z$ is the redshift); a dot
 denotes the derivative with respect to cosmic time $t$. However,
 the integral cannot take the fine structure of $H(z)$ into
 consideration and then lose some important information compiled
 in it~(this point is also noticed in e.g.~\cite{r8}). Therefore,
 it is more rewarding to investigate the observational $H(z)$
 data directly.

 The observational $H(z)$ data we used here are based on differential
 ages of the oldest galaxies~\cite{r9}. In~\cite{r10},
 Jimenez~{\it et~al.} obtained an independent estimate for the
 Hubble constant by the method developed in~\cite{r9}, and used it
 to constrain the equation-of-state parameter~(EoS) of dark energy.
 The Hubble parameter depends on the differential age as a function
 of redshift $z$ in the form
 \be{eq2}
 H(z)=-\frac{1}{1+z}\frac{dz}{dt}\,.
 \ee
 Therefore, a determination of $dz/dt$ directly measures
 $H(z)$~\cite{r11}. By using the differential ages of passively
 evolving galaxies determined from the Gemini Deep Deep Survey
 (GDDS)~\cite{r12} and archival data~\cite{r13}, Simon {\it et al.}
 determined $H(z)$ in the range
 $0\,\lsim\, z\,\lsim\, 1.8$~\cite{r11}. The observational $H(z)$
 data from~\cite{r11} are given in Table~\ref{tab1} and shown
 in Figs.~\ref{fig2}--\ref{fig5}.

\begin{table}[htbp]
\begin{center}
\begin{tabular}{c|lllllllll}\hline\hline
 $z$ &\ 0.09 & 0.17 & 0.27 & 0.40 & 0.88 & 1.30 & 1.43
 & 1.53 & 1.75\\ \hline
 $H(z)\ ({\rm km~s^{-1}\,Mpc^{-1})}$ &\ 69 & 83 & 70
 & 87 & 117 & 168 & 177 & 140 & 202\\ \hline
 $1 \sigma$ uncertainty &\ $\pm 12$ & $\pm 8.3$ & $\pm 14$
 & $\pm 17.4$ & $\pm 23.4$ & $\pm 13.4$ & $\pm 14.2$
 & $\pm 14$ &  $\pm 40.4$\\ \hline\hline
\end{tabular}
\end{center}
\caption{\label{tab1} The observational $H(z)$ data~\cite{r10,r11}
(see~\cite{r14,r15} also).}
\end{table}

These observational $H(z)$ data have been used to constrain the dark
 energy potential and its redshift dependence by Simon {\it et al.}
 in~\cite{r11}. Yi and Zhang used them to constrain the parameters of
 holographic dark energy model in~\cite{r16}. In~\cite{r14}, Samushia
 and Ratra have used these observational $H(z)$ data to constrain
 the parameters of $\Lambda$CDM, XCDM and $\phi$CDM models. Some
 relevant works also include~\cite{r8,r15,r16,r17,r18,r19,r20} for
 examples.

By looking carefully on the observational $H(z)$ data given in
 Table~\ref{tab1} and shown in Figs.~\ref{fig2}--\ref{fig5}, we
 notice that two data points near $z\sim 1.5$ and $0.3$ are very
 special. They deviate from the main trend and dip sharply, especially
 the one near $z\sim 1.5$; the $H(z)$ decreases and then increases
 around them. This hints that the effective EoS crossed $-1$ there.
 In our previous work~\cite{r15}, we have confronted ten cosmological
 models with observational $H(z)$ data, and found that the best models
 have an oscillating feature for both $H(z)$ and effective EoS, with
 the effective EoS crossing $-1$ around redshift $z\sim 1.5$, while
 other non-oscillating dark energy models (e.g. $\Lambda$CDM, XCDM,
 vector-like dark energy etc.) cannot catch the main feature of the
 observational $H(z)$ data.

\begin{center}
\begin{figure}[htbp]
\centering
\includegraphics[width=0.75\textwidth]{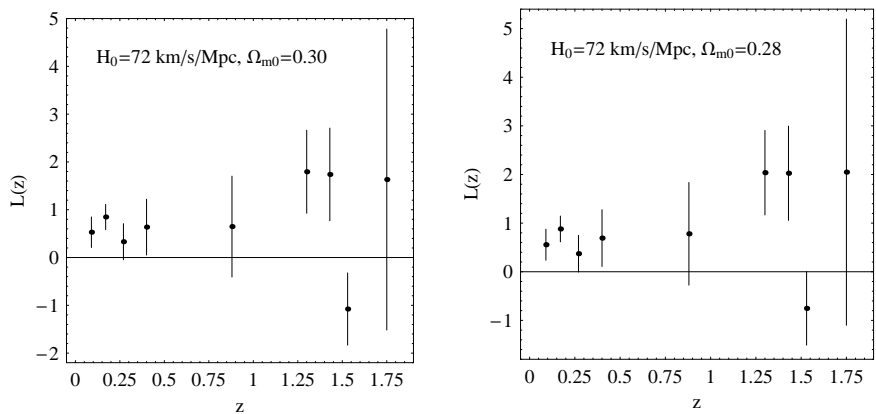}
\caption{\label{fig1} The quantity
 $L(z)\equiv H^2(z)/H_0^2-\Omega_{m0}(1+z)^3$ versus redshift $z$,
 for the fiducial parameters $H_0=72~{\rm km~s^{-1}\,Mpc^{-1}}$ and
 $\Omega_{m0}=0.3$ (left panel) or $0.28$ (right panel).}
\end{figure}
\end{center}

In Fig.~\ref{fig1}, we show the quantity
 $L(z)\equiv H^2(z)/H_0^2-\Omega_{m0}(1+z)^3$ versus redshift $z$,
 which is associated with the fractional energy density of dark
 energy, for the fiducial parameters $H_0=72~{\rm km~s^{-1}\,Mpc^{-1}}$
 and $\Omega_{m0}=0.3$ or $0.28$, where the subscript ``0'' indicates
 the present value of the corresponding quantity. It is easy to see
 that the fractional energy density of dark energy of the point near
 $z\sim 1.5$ is negative (beyond $1\sigma$ significance). To
 avoid this, one can decrease the corresponding $\Omega_{m0}$ or
 make the matter decrease with the expansion of our universe slower
 than $a^{-3}$. Inspired by this, it is natural to consider the
 possibility of exchanging energy between dark energy and dust matter
 through interaction. In fact, we considered the cases with constant
 coupling coefficient in~\cite{r15}. However, we found that it is not
 preferred by the observational $H(z)$ data. In the present work, we
 will explore more forms of couplings between dark energy and dust
 matter, in an attempt to find the couplings which can best describe
 the observational $H(z)$ data.


As extensively considered in the literature (see
 e.g.~\cite{r21,r22,r23,r24,r25,r26,r27,r28,r29,r30,r31,r32,r33,r35,r36,r39,r40,r41,r42,r43}),
 we assume that dark energy and dust matter
 exchange energy through interaction according to
 \bea
 &&\dot{\rho}_X+3H\left(\rho_X+p_X\right)=-3QH\rho_m,\label{eq3}\\
 &&\dot{\rho}_m+3H\rho_m=3QH\rho_m,\label{eq4}
 \eea
 which preserves the total energy conservation equation
 $\dot{\rho}_{tot}+3H\left(\rho_{tot}+p_{tot}\right)=0$. We assume
 that the EoS of dark energy $w_X\equiv p_X/\rho_X$ is constant,
 and consider a spatially flat Friedmann-Robertson-Walker~(FRW)
 universe throughout this work. Notice that the coupling coefficient
 $Q=Q(z)$ can be any function of redshift $z$. So, the interaction
 term $3QH\rho_m$ is a general form, in contrast to the first glance.
 Integrating Eq.~(\ref{eq4}), it is easy to get
 \be{eq5}
 \rho_m\propto\exp\left[\int 3(Q-1)dN\right],
 \ee
 where $N\equiv\ln a=-\ln (1+z)$ is the so-called $e$-folding time;
 the constant proportional coefficient can be determined by
 requiring $\rho_m(N=0)=\rho_{m0}$. Then, $\rho_X$ can be also
 obtained by substituting $\rho_m$ into Eq.~(\ref{eq3}). From the
 Friedmann equation $3H^2=8\pi G(\rho_m+\rho_X)$, the Hubble
 parameter is in hand.

In the following sections, we will compare the observational $H(z)$
 data with some cosmological models with different couplings. We
 adopt the prior $H_0=72~{\rm km~s^{-1}\,Mpc^{-1}}$, which is
 exactly the median value of  the result from the Hubble Space
 Telescope (HST) key project~\cite{r34}, and is also well consistent
 with the one from WMAP 3-year result~\cite{r4}. Since there are
 only 9 observational $H(z)$ data points and their errors are
 fairly large, they cannot severely constrain model parameters
 alone. We perform a $\chi^2$ analysis and compare the cosmological
 models to find out the one which catches the main features of the
 observational $H(z)$ data. We determine the best-fit values for
 the model parameters by minimizing
 \be{eq6}
 \chi^2(parameters)=\sum\limits_{i=1}^9\frac{\left[H_{mod}(parameters;z_i)
 -H_{obs}(z_i)\right]^2}{\sigma^2(z_i)}\,,
 \ee
 where $H_{mod}$ is the predicted value for the Hubble parameter in the
 assumed model, $H_{obs}$ is the observed value, $\sigma$ is the
 corresponding $1\sigma$ uncertainty, and the summation is over the
 9 observational $H(z)$ data points at redshift $z_i$.

\section{Simple couplings}\label{sec2}

We firstly consider the simplest case, $Q=0$, namely there is no
 interaction between dark energy and dust matter. It is easy to find that
 the Hubble parameter is given by~\cite{r15}
 \be{eq7}
 H(z)=H_0\sqrt{\Omega_{m0}(1+z)^3+(1-\Omega_{m0})(1+z)^{3(1+w_X)}}\,,
 \ee
 where $\Omega_{m0}\equiv 8\pi G\rho_{m0}/(3H_0^2)$ is the present
 fractional energy density of dust matter. By minimizing the
 corresponding $\chi^2$, we find that the best-fit values for model
 parameters are $\Omega_{m0}=0.28$ and $w_X=-0.90$, while
 $\chi^2_{min}=9.02$ for 7 degrees of freedom and
 $P\left(\chi^2>\chi^2_{min}\right)=0.25$.

Next, we consider the case of $Q=const.$. By using Eqs.~(\ref{eq5})
 and~(\ref{eq3}), we can obtain the Hubble parameter as~\cite{r15}
 \be{eq8}
 H(z)=H_0\sqrt{\frac{w_X \Omega_{m0}}{Q+w_X}(1+z)^{3(1-Q)}
 +\left(1-\frac{w_X \Omega_{m0}}{Q+w_X}\right)(1+z)^{3(1+w_X)}}\,.
 \ee
 By minimizing the corresponding $\chi^2$, we find that the best-fit
 values for model parameters are $\Omega_{m0}=0.72$, $w_X=-3.70$ and
 $Q=0.30$, while $\chi^2_{min}=8.48$ for 6 degrees of freedom and
 $P\left(\chi^2>\chi^2_{min}\right)=0.21$. Although the
 $\chi^2_{min}$ is reduced slightly compare to the case of $Q=0$,
 statistically this fitting has no significant improvement over the
 case of $Q=0$, since the number of parameters is increased from 2 to 3.

Another $Q$ with only one parameter comes from the
 assumption~\cite{r33,r35,r36}
 \be{eq9}
 \frac{\rho_X}{\rho_m}=\frac{\rho_{X0}}{\rho_{m0}}a^\xi,
 \ee
 where $\xi$ is a constant parameter, which quantifies the severity
 of the coincidence problem. Following~\cite{r33,r35,r36}, it is
 easy to find that the corresponding $Q$ is given by (which is
 labeled as SCL)
 \be{eq10}
 Q(z)=\frac{Q_0}{1-\Omega_{m0}+\Omega_{m0}(1+z)^\xi},
 \ee
 where $Q_0=-(1-\Omega_{m0})(\xi+3w_X)/3$; and the Hubble parameter
 reads
 \be{eq11}
 H(z)=H_0(1+z)^{3/2}\left[\Omega_{m0}+
 (1-\Omega_{m0})(1+z)^{-\xi}\right]^{-3w_X/(2\xi)}.
 \ee
 By minimizing the corresponding $\chi^2$, we find that the best-fit
 values for model parameters are $\Omega_{m0}=0.92$, $w_X=-8.72$ and
 $\xi=1.49$, while $\chi^2_{min}=8.88$ for 6 degrees of freedom and
 $P\left(\chi^2>\chi^2_{min}\right)=0.18$. It is worth noting that
 the best-fit value $\Omega_{m0}=0.92$ is inconsistent with the
 results from clusters of galaxies~\cite{r38} and 3-year
 WMAP~\cite{r4} etc.

\begin{center}
\begin{figure}[htbp]
\centering
\includegraphics[width=0.75\textwidth]{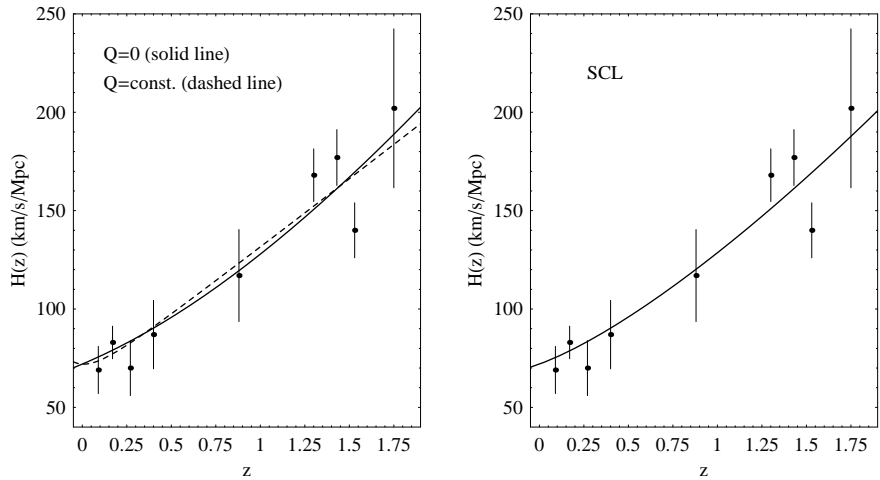}
\caption{\label{fig2} The observational $H(z)$ data with error bars,
 and the theoretical lines for three simple models with the
 corresponding best-fit parameters. Left panel: the cases of
 $Q=0$~(solid line) and $Q=const.$~(dashed line); Right panel: the
 case of SCL.}
\end{figure}
\end{center}

We present the observational $H(z)$ data with error bars, and the
 theoretical lines for these simple models with the corresponding
 best-fit parameters in Fig.~\ref{fig2}. In fact, the cases of
 $Q=0$ and SCL cannot be distinguished significantly. It can be
 seen clearly from Fig.~\ref{fig2} that none of them may reproduce
 the sharp dip around $z\sim 1.5$; the data point near $z\sim 1.5$
 deviates from model fitting by about $2\sigma$. In fact, it can
 be seen from Table~\ref{tab2}, all of these models are statistically
 even worse than the $\Lambda$CDM model.


\section{More complicated couplings}\label{sec3}

In this section, we will consider some more complicated couplings.
 Unlike in Section~\ref{sec2}, for these complicated cases, it is
 difficult to obtain the analytical $\rho_X(z)$ from Eq.~(\ref{eq3})
 with Eq.~(\ref{eq5}), and then the Hubble parameter $H(z)$.
 Therefore, the numerical methods~\cite{r37} are necessary. For the
 convenience of numerical computing, we introduce
 $\tilde{\rho}_m\equiv 8\pi G\rho_m/(3H_0^2)$ and
 $\tilde{\rho}_X\equiv 8\pi G\rho_X/(3H_0^2)$. Eqs.~(\ref{eq3})
 and~(\ref{eq5}) become
 \bea
 &&\frac{d\tilde{\rho}_X}{dN}=-3\tilde{\rho}_X
 (1+w_X)-3Q\tilde{\rho}_m,\label{eq12}\\
 &&\tilde{\rho}_m\propto\exp\left[\int 3(Q-1)dN\right],\label{eq13}
 \eea
 where the constant proportional coefficient in Eq.~(\ref{eq13}) is
 determined by requiring $\tilde{\rho}_m(N=0)=\Omega_{m0}$; the
 initial condition for integrating Eq.~(\ref{eq12}) is
 $\tilde{\rho}_X(N=0)=1-\Omega_{m0}$. The Hubble parameter is given
 by $H=H_0(\tilde{\rho}_X+\tilde{\rho}_m)^{1/2}$. In the following
 subsections, we will consider some cases with {\em parameterized} $Q(N)$.
 We restrict ourselves to the cases with only two parameters, since
 the data points are so few. To find the minimal $\chi^2$ efficiently,
 we scan the parameters space with a relatively large grid size at
 first, and then narrow the parameters space and scan with a small
 grid size. In the procedure, we impose the condition
 $\tilde{\rho}_X\ge 0$, while $\tilde{\rho}_m\ge 0$ has been
 guaranteed by Eq.~(\ref{eq13}) for real $Q(N)$.

\subsection{Linear coupling}\label{sec3a}

Here, we consider a linear coupling~(LIN)
 \be{eq14}
 Q(N)=Q_0+Q_1 N,
 \ee
 where $Q_0$ and $Q_1$ are constants. The corresponding
 $\tilde{\rho}_m$ is given by
 \be{eq15}
 \tilde{\rho}_m=\Omega_{m0}\exp\left[3(Q_0-1)N
 +\frac{3}{2}Q_1 N^2\right].
 \ee
 We find that the best-fit parameters are $\Omega_{m0}=0.001$,
 $w_X=-0.80$, $Q_0=-7.78$ and $Q_1=-14.32$, while
 $\chi^2_{min}=7.08$ for 5 degrees of freedom and
 $P\left(\chi^2>\chi^2_{min}\right)=0.21$.


\begin{center}
\begin{figure}[htbp]
\centering
\includegraphics[width=0.75\textwidth]{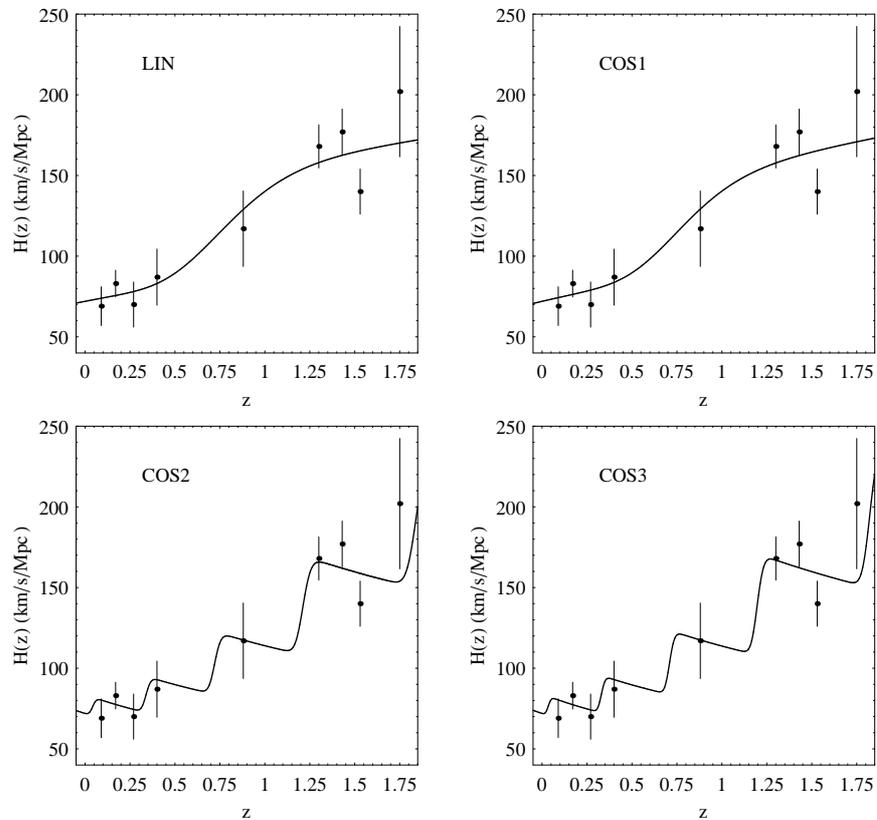}
\caption{\label{fig3} The observational $H(z)$ data with error bars,
 and the theoretical lines for the cases of LIN, COS1, COS2 and COS3,
 with the corresponding best-fit parameters.}
\end{figure}
\end{center}


\subsection{Couplings of cosine type}\label{sec3b}

We consider a coupling of cosine type~(COS1) as
 \be{eq16}
 Q(N)=A_1\cos(A_2 N),
 \ee
 where $A_1$ and $A_2$ are constants. The corresponding
 $\tilde{\rho}_m$ is given by
 \be{eq17}
 \tilde{\rho}_m=\Omega_{m0}\exp\left[-3N
 +3\frac{A_1}{A_2}\sin\left(A_2 N\right)\right].
 \ee
 We find that the best-fit parameters are $\Omega_{m0}=0.001$,
 $w_X=-0.76$, $A_1=-6.09$ and $A_2=2.87$, while
 $\chi^2_{min}=7.07$ for 5 degrees of freedom and
 $P\left(\chi^2>\chi^2_{min}\right)=0.22$.

The next one~(COS2) is given by
 \be{eq18}
 Q(N)=A_1^3\cos\left(N/A_2+A_1\pi\right).
 \ee
 It is easy to find that
 \be{eq19}
 \tilde{\rho}_m=\Omega_{m0}\exp\left\{-3N+3A_1^3A_2
 \left[\sin\left(N/A_2+A_1\pi\right)
 -\sin\left(A_1\pi\right)\right]\right\}.
 \ee
 We find that the best-fit parameters are $\Omega_{m0}=0.07$,
 $w_X=-1.33$, $A_1=-3.83$ and $A_2=-0.04$, while
 $\chi^2_{min}=5.89$ for 5 degrees of freedom and
 $P\left(\chi^2>\chi^2_{min}\right)=0.32$.

We consider the third case~(COS3)
 \be{eq20}
 Q(N)=A_1^3\cos\left(N/A_2+\pi/A_1\right).
 \ee
 The corresponding $\tilde{\rho}_m$ reads
 \be{eq21}
 \tilde{\rho}_m=\Omega_{m0}\exp\left\{-3N+3A_1^3A_2
 \left[\sin\left(N/A_2+\pi/A_1\right)
 -\sin\left(\pi/A_1\right)\right]\right\}.
 \ee
 The best-fit parameters are found to be $\Omega_{m0}=0.07$,
 $w_X=-1.35$, $A_1=-4.41$ and $A_2=0.04$, while
 $\chi^2_{min}=5.88$ for 5 degrees of freedom and
 $P\left(\chi^2>\chi^2_{min}\right)=0.32$.

In Fig.~\ref{fig3}, we present the observational $H(z)$ data with
 error bars, and the theoretical lines for the cases of LIN, COS1,
 COS2 and COS3, with the corresponding best-fit parameters. It
 can be seen clearly from Fig.~\ref{fig3} that the cases of LIN
 and COS1 cannot reproduce the sharp dip around $z\sim 1.5$.
 Although the cases of COS3 and COS4 have the oscillating feature,
 the data points near $z\sim 1.5$ and $1.75$ deviate from model
 fitting beyond $1\sigma$. We notice that the best-fit parameter
 $\Omega_{m0}$ for the cases of LIN, COS1, COS2 and COS3 are too
 small to be consistent with the results from clusters of
 galaxies~\cite{r38} and 3-year WMAP~\cite{r4} etc. After all, even
 purely from statistical point of view, all these models are not
 preferred over the $\Lambda$CDM model.

\subsection{Couplings of Gaussian distribution type}\label{sec3c}

In this subsection, we consider some couplings of Gaussian
 distribution type. In fact, they can efficiently mimic the
 $\delta$ function, namely, these $Q$ can be rather large
 in a very narrow range of $N$ and are approximately zero
 for other $N$.

We consider the case of original Gaussian distribution~(GD1)
 at first,
 \be{eq22}
 Q(N)=\frac{1}{\sqrt{\pi}\,Q_s}\exp
 \left[-\frac{(N-N_s)^2}{Q_s^2}\right],
 \ee
 where $Q_s$ and $N_s$ are constants. The corresponding
 $\tilde{\rho}_m$ is given by
 \be{eq23}
 \tilde{\rho}_m=\Omega_{m0}\exp\left[-3N
 +\frac{3}{2}\,{\rm Erf}\left(\frac{N-N_s}{Q_s}\right)
 +\frac{3}{2}\,{\rm Erf}\left(\frac{N_s}{Q_s}\right)\right],
 \ee
 where $\disp {\rm Erf}(x)\equiv\frac{2}{\sqrt{\pi}}\int_0^x
 e^{-t^2}dt$ is the well-known error function~\cite{r37}.
 We find that the best-fit parameters are $\Omega_{m0}=0.43$,
 $w_X=-1.64$, $Q_s=0.007$ and $N_s=-0.85$, while
 $\chi^2_{min}=5.89$ for 5 degrees of freedom and
 $P\left(\chi^2>\chi^2_{min}\right)=0.32$.

Then, we consider a variant~(GD2) of the original Gaussian
 distribution as
 \be{eq24}
 Q(N)=\frac{2}{\sqrt{\pi}\,Q_s^3}\exp
 \left[-\frac{(N-N_s)^2}{Q_s^2}\right],
 \ee
 and find that
 \be{eq25}
 \tilde{\rho}_m=\Omega_{m0}\exp\left\{-3N
 +\frac{3}{Q_s^2}\left[{\rm Erf}\left(\frac{N-N_s}{Q_s}\right)
 +{\rm Erf}\left(\frac{N_s}{Q_s}\right)\right]\right\}.
 \ee
 We obtain the best-fit parameters as $\Omega_{m0}=0.43$,
 $w_X=-1.61$, $Q_s=0.014$ and $N_s=-0.89$, while
 $\chi^2_{min}=5.93$ for 5 degrees of freedom and
 $P\left(\chi^2>\chi^2_{min}\right)=0.31$.


\begin{center}
\begin{figure}[htbp]
\centering
\includegraphics[width=0.75\textwidth]{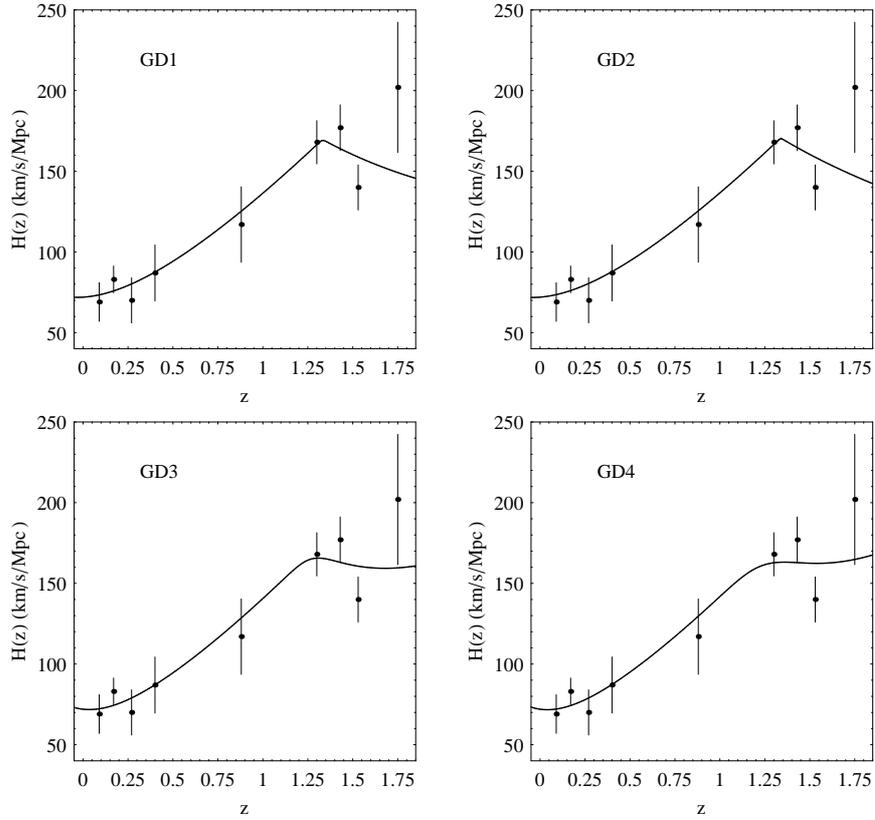}
\caption{\label{fig4} The observational $H(z)$ data with error bars,
 and the theoretical lines for the cases of GD1, GD2, GD3 and GD4,
 with the corresponding best-fit parameters.}
\end{figure}
\end{center}


Also, other variant~(GD3) of the original Gaussian distribution
 can be the form of
 \be{eq26}
 Q(N)=\frac{2}{\sqrt{\pi}\,Q_s}\exp
 \left[-\frac{(N-N_s)^2}{Q_s^4}\right].
 \ee
 Its corresponding $\tilde{\rho}_m$ is given by
 \be{eq27}
 \tilde{\rho}_m=\Omega_{m0}\exp\left\{-3N
 +3Q_s\left[{\rm Erf}\left(\frac{N-N_s}{Q_s^2}\right)
 +{\rm Erf}\left(\frac{N_s}{Q_s^2}\right)\right]\right\}.
 \ee
 We find that the best-fit parameters are $\Omega_{m0}=0.47$,
 $w_X=-2.10$, $Q_s=0.21$ and $N_s=-0.83$, while
 $\chi^2_{min}=6.07$ for 5 degrees of freedom and
 $P\left(\chi^2>\chi^2_{min}\right)=0.30$.

We consider the last one~(GD4) as
 \be{eq28}
 Q(N)=\frac{2}{\sqrt{\pi}\,Q_s}\exp
 \left[-\frac{(N-N_s)^2}{Q_s^6}\right].
 \ee
 It is easy to find that
 \be{eq29}
 \tilde{\rho}_m=\Omega_{m0}\exp\left\{-3N
 +3Q_s^2\left[{\rm Erf}\left(\frac{N-N_s}{Q_s^3}\right)
 +{\rm Erf}\left(\frac{N_s}{Q_s^3}\right)\right]\right\}.
 \ee
 We obtain the best-fit parameters as $\Omega_{m0}=0.48$,
 $w_X=-2.20$, $Q_s=0.42$ and $N_s=-0.82$, while
 $\chi^2_{min}=6.43$ for 5 degrees of freedom and
 $P\left(\chi^2>\chi^2_{min}\right)=0.27$.

In Fig.~\ref{fig4}, we present the observational $H(z)$ data with
 error bars, and the theoretical lines for the cases of GD1, GD2,
 GD3 and GD4, with the corresponding best-fit parameters. It
 can be seen clearly from Fig.~\ref{fig4} that the data points
 near $z\sim 1.5$ and $1.75$ deviate from model fitting about
 $1\sigma$. Again, none of these models is better than the
 $\Lambda$CDM model.

\section{Conclusion and discussions}\label{sec4}

In this note, we extend our previous work~\cite{r15}, and compare
 eleven interacting dark energy models with different couplings
 to the observational $H(z)$ data. In Table~\ref{tab2}, we summarize
 all eleven models considered in this work. In addition, we show the
 results for the simplest $\Lambda$CDM model from~\cite{r15} together.
 Obviously, although the $\chi^2_{min}$ of all interacting dark energy
 models are lower than the simplest $\Lambda$CDM model, their
 $\chi^2_{min}/dof$ are higher, and their
 $P\left(\chi^2>\chi^2_{min}\right)$ are lower correspondingly.
 This implies that either more exotic couplings are needed in the
 cosmological models with interaction between dark energy and dust
 matter, or {\em there is no interaction at all}. We consider that
 this result is disadvantageous to the interacting dark energy models
 studied extensively in the literature (see~\cite{r21,r22,r23,r24,
 r25,r26,r27,r28,r29,r30,r31,r32,r33,r35,r36,r39,r40,r41,r42,r43}
 for examples).

Although the simplest $\Lambda$CDM model looks better, however, it
 is not the best model which is preferred by the observational
 $H(z)$ data. In fact, as shown in our previous work~\cite{r15},
 the observational $H(z)$ data favors the models which have an
 oscillating feature for both $H(z)$ and effective EoS, with the
 effective EoS crossing $-1$ around redshift $z\sim 1.5$. Since as
 shown in the present work the interacting dark energy models fail,
 other physical mechanisms are needed to produce an oscillating
 feature for $H(z)$ which can also fit the observational $H(z)$ data
 fairly well. We leave this to our future works.

We stress that other current data, such as SNe~Ia~\cite{r2,r3,r7},
 CMB~\cite{r4} and so on, are not inconsistent with the
 $\Lambda$CDM model, as shown in e.g.~\cite{r44}. Even for the
 observational $H(z)$ data, we notice that from Table~\ref{tab2},
 the $\Lambda$CDM model has $\chi^2_{min}=9.04$ for
 eight degrees of freedom with $34\%$ probability, which is not
 unacceptably low. Before the new and improved $H(z)$ data are
 available, it is too early to say that the $\Lambda$CDM model
 can be ruled out.

\begin{table}[htbp]
\begin{center}
\begin{tabular}{c|c|c|c}\hline\hline
 Model & $\hspace{4mm}\chi_{min}^2\hspace{4mm}$
 & $\ \chi_{min}^2/dof\ $
 & $P\left(\chi^2>\chi_{min}^2\right)$\\ \hline
 $\Lambda$CDM & 9.04 & 1.13 & 0.34\\
 $Q=0$ & 9.02 & 1.29 & 0.25\\
 $Q=const.$ & 8.48 & 1.41 & 0.21\\
 SCL & 8.88 & 1.48 & 0.18\\
 LIN & 7.08 & 1.42 & 0.21\\
 COS1 & 7.07 & 1.41 & 0.22\\
 COS2 & 5.89 & 1.18 & 0.32\\
 COS3 & 5.88 & 1.18 & 0.32\\
 GD1 & 5.89 & 1.18 & 0.32\\
 GD2 & 5.93 & 1.19 & 0.31\\
 GD3 & 6.07 & 1.21 & 0.30\\
 GD4 & 6.43 & 1.29 & 0.27\\
 \hline\hline
\end{tabular}
\end{center}
\caption{\label{tab2} Summarizing all eleven models considered
 in this work. In addition, we show the results for the simplest
 $\Lambda$CDM model from~\cite{r15} together. We have adopted
 $H_0=72~{\rm km~s^{-1}\,Mpc^{-1}}$ exactly in the computations.}
\end{table}

In fact, we can also consider the couplings with three parameters,
 such as $Q(N)=A_1\cos (A_2 N+A_3\pi)$,
 $Q(N)=A_s\exp\left[-(N-N_s)^2/Q_s^2\right]$,
 $Q(N)=A_s\exp\left[-(N-N_s)^2/Q_s^4\right]$ and so on.
 However, they are less attractive, since the data points
 are so few. Also, in addition to the couplings of Gaussian
 distribution type, $Q(N)$ can mimic $\delta$ function
 through e.g. $Q(N)=Q_h\,{\rm sech}[2(N-N_h)]$,
 $Q(N)=Q_h\,{\rm sech}[2Q_h(N-N_h)]$,
 $Q(N)=Q_h\,{\rm sech}[2Q_h^2(N-N_h)]$ and so on, where
 ${\rm sech}(x)=1/\cosh(x)$. But we do not consider these
 cases in this work any more, since it is expected that
 they are physically similar to the cases of Gaussian
 distribution type.

If we discard the condition $\tilde{\rho}_X\ge 0$, the fit
 can be improved, such as $\chi^2_{min}=4.63$,
 $\chi^2_{min}/dof=0.93$ and
 $P\left(\chi^2>\chi^2_{min}\right)=0.46$
 for parameters $\Omega_{m0}=0.02$,
 $w_X=-1.73$, $A_1=3.16$ and $A_2=0.03$ for the case of COS2,
 while  $\chi^2_{min}=5.77$, $\chi^2_{min}/dof=1.15$ and
 $P\left(\chi^2>\chi^2_{min}\right)=0.33$ for
 parameters $\Omega_{m0}=0.57$, $w_X=-2.41$, $A_1=1.35$
 and $A_2=0.09$ for the case of COS3. However, we consider
 that they are not physically acceptable, since their
 $\tilde{\rho}_X$ become negative for some intervals of
 redshift $z$.

\begin{center}
\begin{figure}[htbp]
\centering
\includegraphics[width=0.75\textwidth]{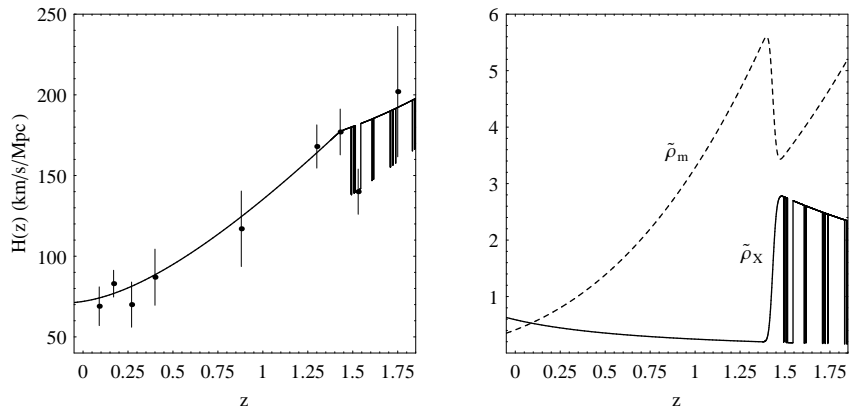}
\caption{\label{fig5} The numerical results for the case of GD3
 with the corresponding ``best-fit'' parameters, which come from
 a careless numerical computing. See text for details.}
\end{figure}
\end{center}

It is worth noting that for the cases with couplings of
 Gaussian distribution type, careful treatment in the
 numerical computing is necessary. If one naively uses
 the Bulirsch-Stoer method or the adaptive Runge-Kutta
 method~\cite{r37} to integrate Eq.~(\ref{eq12}) directly,
 misleading results can be obtained, mainly due to the
 $\delta$ function feature of the Gaussian distribution.
 For instance, one may find that $\chi^2_{min}=1.64$,
 $\chi^2_{min}/dof=0.33$ and
 $P\left(\chi^2>\chi^2_{min}\right)=0.90$ for parameters
 $\Omega_{m0}=0.41$, $w_X=-1.42$, $Q_s=0.10$
 and $N_s=-0.89$ for the case of GD3. However, this is
 misleading. We present the corresponding results of $H(z)$
 and $\tilde{\rho}_X$, $\tilde{\rho}_m$ in Fig.~\ref{fig5}.
 Obviously, the unaccounted fiber-like dips are not
 physical results. By using a more delicate treatment in
 the numerical computing, the right results can be obtained,
 as presented in Section~\ref{sec3c} and Fig.~\ref{fig4}.

\begin{table}[htbp]
\begin{center}
\begin{tabular}{c|c|c|c}\hline\hline
 Model & $\hspace{4mm}\chi_{min}^2\hspace{4mm}$
 & $\ \chi_{min}^2/dof\ $
 & $P\left(\chi^2>\chi_{min}^2\right)$\\ \hline
 $\Lambda$CDM & 9.03 & 1.29 & 0.25\\
 $Q=0$ & 8.88 & 1.48 & 0.18\\
 $Q=const.$ & 8.48 & 1.70 & 0.13\\
 SCL & 8.81 & 1.76 & 0.12\\
 \hline\hline
\end{tabular}
\end{center}
\caption{\label{tab3} Summarizing the four models in which
 $H_0$ is considered as a free parameter with the prior
 $H_0=72\pm 8~{\rm km~s^{-1}\,Mpc^{-1}}$ at $1\,\sigma$
 uncertainty~\cite{r34}.}
\end{table}

After all, it is worth noting that we adopt the exact
 $H_0=72~{\rm km~s^{-1}\,Mpc^{-1}}$ throughout this work.
 However, we admit that the uncertainty in $H_0$ should be
 taken into account. For instance, it is suggested that
 in~\cite{r34}
 $H_0=72\pm 8~{\rm km~s^{-1}\,Mpc^{-1}}$ at $1\,\sigma$
 uncertainty, in~\cite{r45}
 $H_0=68\pm 7~{\rm km~s^{-1}\,Mpc^{-1}}$ at $2\,\sigma$
 uncertainty, and in~\cite{r46}
 $H_0=62.3\pm 6.3~{\rm km~s^{-1}\,Mpc^{-1}}$ at $1\,\sigma$
 uncertainty. So, we should examine the effects of the uncertainty
 in $H_0$ on our conclusions. Here we consider the four simplest
 models studied in this work for examples. They are the models
 $\Lambda$CDM, $Q=0$, $Q=const.$ and SCL. The $H_0$ is
 considered as a free parameter in these models, with the
 prior $H_0=72\pm 8~{\rm km~s^{-1}\,Mpc^{-1}}$ at $1\,\sigma$
 uncertainty~\cite{r34}. In this case, the total $\chi^2$
 should be the summation of the one in Eq.~(\ref{eq6}) and
 $\chi^2_{H0}=(H_0-72)^2/8^2$. We find that the best-fit values for
 model parameters of $\Lambda$CDM are $\Omega_{m0}=0.31$ and
 $H_0=71.44~{\rm km~s^{-1}\,Mpc^{-1}}$, while $\chi^2_{min}=9.03$
 for 7 degrees of freedom and
 $P\left(\chi^2>\chi^2_{min}\right)=0.25$. For model $Q=0$, the
 best-fit parameters are $\Omega_{m0}=0.18$, $w_X=-0.45$ and
 $H_0=68.53~{\rm km~s^{-1}\,Mpc^{-1}}$, while $\chi^2_{min}=8.88$
 for 6 degrees of freedom and
 $P\left(\chi^2>\chi^2_{min}\right)=0.18$. For model $Q=const.$,
 the best-fit parameters are $\Omega_{m0}=0.72$, $w_X=-3.68$,
 $Q=0.30$ and $H_0=71.88~{\rm km~s^{-1}\,Mpc^{-1}}$,
 while $\chi^2_{min}=8.48$ for 5 degrees of freedom and
 $P\left(\chi^2>\chi^2_{min}\right)=0.13$. For model SCL, the
 best-fit parameters are $\Omega_{m0}=0.94$, $w_X=-8.45$,
 $\xi=0.88$ and $H_0=69.79~{\rm km~s^{-1}\,Mpc^{-1}}$, while
 $\chi^2_{min}=8.81$ for 5 degrees of freedom and
 $P\left(\chi^2>\chi^2_{min}\right)=0.12$. We summarize these
 results in Table~\ref{tab3}. Comparing with the corresponding
 results of these four models in Table~\ref{tab2} where $H_0$ is
 taken as $72~{\rm km~s^{-1}\,Mpc^{-1}}$ exactly, we find that
 the uncertainty in $H_0$ cannot significantly improve the fits.
 Thus, even the uncertainty in $H_0$ is taken into
 consideration, our conclusions remain unchanged.


\section*{Acknowledgments}
We thank the anonymous referees for quite useful comments and
 suggestions, which help us to improve this work. We are grateful to
 Prof.~Rong-Gen~Cai and Prof.~Zong-Hong~Zhu for helpful
 discussions. We also thank Zong-Kuan~Guo, Hongsheng~Zhang, Xin~Zhang,
 Hui~Li, Meng~Su and Ningning~Tang, Sumin~Tang,
 Shi-Chao~Tang, Jian~Hu, Yue~Shen, Xin~Liu, Lin~Lin, Jing~Jin,
 Wei-Ke~Xiao, Feng-Yun~Rao, Nan~Liang, Rong-Jia~Yang,
 Jian~Wang, Yuan~Liu for kind help and discussions. We
 acknowledge partial funding support from China Postdoctoral Science
 Foundation, and by the Ministry of Education of China, Directional
 Research Project of the Chinese Academy of Sciences under project
 No.~KJCX2-YW-T03, and by the National Natural Science Foundation
 of China under project No.~10521001.


\end{document}